\documentclass[manuscript,screen,anonymous=false,nonacm]{acmart}

\AtBeginDocument{%
  \providecommand\BibTeX{{%
    \normalfont B\kern-0.5em{\scshape i\kern-0.25em b}\kern-0.8em\TeX}}}

\setcopyright{acmcopyright}
\copyrightyear{2021}
\acmYear{2021}
\acmDOI{}

\acmConference[RecSys '21]{RecSys '21: Fifteenth ACM Conference on Recommender Systems}{27 September -- 01 October, 2021}{Amsterdam, Netherlands}
\acmPrice{}
\acmISBN{}

\usepackage[ruled,vlined]{algorithm2e}
\usepackage[skip=5pt]{caption}

\begin{document}

\title{Automatic Collection Creation and Recommendation}

\author{Sanidhya Singal}
\authornote{Both authors contributed equally to this research.}
\email{sanidhya.singal@wynk.in}
\affiliation{%
  \institution{Airtel Digital}
  \city{Gurugram}
  \country{India}
}
\author{Piyush Singh}
\authornotemark[1]
\email{piyush.singh@wynk.in}
\affiliation{%
  \institution{Airtel Digital}
  \city{Gurugram}
  \country{India}
}
\author{Manjeet Dahiya}
\email{manjeet.dahiya@wynk.in}
\affiliation{%
  \institution{Airtel Digital}
  \city{Gurugram}
  \country{India}
}


\begin{abstract}
We present a \emph{collection recommender system} that can automatically
create and recommend collections of items at a user level.
Unlike regular recommender systems, which output
top-N relevant items, a collection recommender system
outputs collections of items such that the items in the collections 
are relevant to a user, 
and the items within a collection follow a specific theme.
Our system builds on top of the user-item representations learnt by 
item recommender systems.
We employ dimensionality reduction and clustering techniques 
along with intuitive heuristics 
to create collections with their ratings and titles.

We test these ideas in a real-world setting of music recommendation, 
within a popular music streaming service.
We find that there is a 2.3x increase in recommendation-driven consumption
when recommending collections over items.
Further, it results in effective utilization of
real estate and leads to recommending a more and diverse set of items.
To our knowledge, these are first of its kind experiments at such a large scale.

\end{abstract}


\maketitle

\section{Introduction}

In this paper, we present a \emph{collection recommender system} that can automatically
create and recommend collections of items.
A regular recommender system \cite{als-paper, als-mf-paper, nonneg-mf-paper, als-wr-paper, amazon-item-cf-paper, neural-cf-paper, youtube-dl-recsys-paper, deep-learning-recsys-paper, recsys-dl-google} usually outputs top-N items 
that are relevant to a user.
On the other hand, a collection recommender system
outputs collections of items such that the items in the collections are relevant to a user, 
and within a collection, the items follow a specific theme of belonging.

A concrete instance of a collection recommender system is music recommendation, where
recommending playlists (i.e., collections) is more desirable 
than recommending individual songs (i.e., items).
A typical recommendation of top-N relevant songs
results in the songs spanning
various themes, which is unsuitable for consumption.
For example, users would not want to listen to a heavy metal song
after a soothing western classical.
The ideal recommendation in this setting is a handful of playlists
whose content is relevant to users,
and at the same time, songs within a playlist are similar.
A user can conveniently select a collection  
as per his/her current mood and interest,
and then consume the entire collection without abrupt theme changes.

Collection recommendation aptly fits the domain of music, and
it has wider applications.
For example, in movie recommendation, we need to recommend personalized collections based on
specific genres such as thriller and comedy 
\cite{netflix-collections-paper, netflix-business-paper}.
In case of e-commerce, collections are recommended based on the product categories 
\cite{amazon-recsys-paper}. 
Search and information retrieval is another area where collection based presentation 
is relatively a better approach.
Here as well, we can organize the search results in collections and a user can then
dive into the desired collection.
Hansen and Golbeck \cite{hansen-golbeck-mix-paper} discuss more on 
the importance of collection recommendation from 
the perspective of usability and present several use cases.

The problem of collection recommendation is not prominent in e-commerce and movies recommendation
because various tags of items (e.g., category and genre information) are readily available. 
With the availability of these tags, creating collections is straightforward 
--- we just need to filter the top-N recommendations by 
tags to create the respective collections.
In contrast, in a setting where tags are not available, creating collections is non-trivial.
Music is one such example where tags are either unavailable or 
not comprehensive enough.
Similarly, even in the setting of video (user-generated content) recommendation, 
we may not have a comprehensive tagging, 
e.g., 
``festival'' and  
``home makeover'' are unique tags, which might be unavailable.
In such settings, a collection recommender system can output collections of varying themes.

We present the design of a collection recommender system, which is capable of automatically
creating and
recommending collections without the need for tagged items.
Thus, our system targets a relatively general setting and would work in most scenarios.
We build it on top of regular recommender systems that are used for recommending top-N
items. Specifically, our system uses the user-item representations
learnt through the training of an item recommender system.
The latter could be a matrix factorization based collaborative filtering 
recommender system such as ALS \cite{als-paper, als-mf-paper, als-wr-paper}, 
or
it could very well be a deep neural network based recommender system \cite{neural-cf-paper, youtube-dl-recsys-paper, deep-learning-recsys-paper, recsys-dl-google}.
The only requirement is that the underlying model exposes 
user-item representations (embeddings).

Our system uses these embeddings to generate top-N recommendations
for each user and then clusters them.
We perform clustering on the 
item embeddings after reducing their dimensionality.
We then form collections by picking the relevant and representative 
items in each cluster.
The basis for this to work is that the learnt embeddings provide 
the notion of similarity between items as well as the ability to predict
the relevance of an item to a user (more in Section~\ref{solution}). 
Finally, we compute the ratings and the representative titles of the clusters
after few hygiene checks.

We test our ideas in a real-world setting within a music streaming service.
We qualitatively show that collections created with our system are rich 
and personalized to a user.
On manually inspecting the collections, we found that 
our system has been able to create nuanced collections with themes such as 
``religious'', 
``movie themes'',
and ``pop songs of the 90s''.
Annotating tags of such nature is hard and laborious especially when 
labelling items individually.
Moreover, new items keep coming in the catalogue and manual labelling
is impractical.
Finally, we quantitatively report that the collection based recommendation approach can be 
extremely effective and lead to a 2.3x increase in recommendation-driven consumption 
(more in Section~\ref{online-experiments}) over the
approach of item recommendation.
Further, it results in effective utilization of
real estate and leads to recommending a more and diverse set of items.

Our work makes the following contributions:
1) We present the design of a system that can create and recommend collections
of items. The system does not require additional training data, and
it works with the same data that is needed to
train item recommender systems.
2) We present comparative results between item and
collection recommendation in a real-world setting of
music recommendation. To our knowledge, 
these are first of its kind experiments at such a large scale.

\section{Problem Formulation}

We are given a set of items $I$ and a set of users $U$ along with 
the interaction data $\{(u, i, r_{ui})\}$, where
$i \in I$, $u \in U$ and $r_{ui}$ represents the implicit or explicit rating 
derived from the historical consumption of 
item $i$ by user $u$.
Additionally, we can also have side information such as user attributes, 
item attributes and contextual information depending upon the scope of 
the item recommender system that we use to learn the representations.

Using this data, our goal is to recommend up to $N$ collections 
$C^u = [C_1^u, C_2^u, ..., C_N^u]$ to user $u$. 
A collection $C_k^u, k \in [1, N]$, is defined as 
$(\text{title}_k^u, R_k^u, \{(i_k^u, r_k^u): i_k^u \in I\})$, 
where $\text{title}_k^u$ is a textual representation of the collection, 
which can be shown to the user; 
$R_k^u$ is the overall rating of the collection with respect to user $u$; 
$i_k^u$ refers to an item present in the collection, 
and $r_k^u$ denotes the rating of the respective item for the  
user $u$ as predicted by the recommender system. 
Each collection contains a fixed number of items.

The desired properties of the collections are as follows:
1) The two kinds of ratings predicted ($R_k^u$ and $r_k^u$)
should be such that they ensure that
the collection, as well as the items within a collection, are relevant to the
user.
2) Two different items within a collection should be similar to each other based 
on a given similarity criteria.
3) Two collections should be different from each other so that
a diverse set of collections is recommended.
4) The title of the collection should be representative of the items
present in the collection.

\section{Solution} \label{solution}

Our system builds on top of the capability of existing recommender systems of learning
user and item representations.
We first train an item recommender system to learn these representations and then
use them to create and recommend collections.

\subsection{Learning user and item representation}
We train an item recommender system on 
user-item interaction data $\{ (u, i, r_{ui}) \}$ derived from historical consumption.
Our goal of training a recommender system is 
to learn the representations of users and items.
These representations are typically exposed 
as embeddings 
(or the activations of the penultimate layers in case of deep networks),
which are mathematically 
represented as low dimensional vectors.
These embeddings of users and items are mapped to the same vector space and are 
trained to predict the rating/similarity ($r_{ui}$) between a user and an item. 

Formally, the vectors for an item $i$ and a user $u$ are represented
as $i^{Emb}$ and $u^{Emb}$, respectively.
We train the recommender system to learn embeddings by minimizing the difference between
$r_{ui}$ and $(u^{Emb} \cdot i^{Emb})$, where $\cdot$ denotes the dot product.
The result of the training is that we learn representations of items and users 
that can be used to compute the ratings between them
by taking dot products of the respective embeddings.
More importantly, two different items that have the same behaviour in the training data, 
i.e., similar user-item interaction (or otherwise consumed by the same set of users), 
would result in their embeddings that are close to each other in the Euclidean space.
It implies that the representations are generalized to 
predict item-to-item similarity also \cite{item-to-item-sim-paper}.

We use these notions of similarities to construct collections of similar items 
and determine their relevance ($R^u$) to a user.
Thus, we require a recommender system that exposes these embeddings.  
Most of the common recommender systems \cite{basic-recsys-paper, als-wr-paper, deep-learning-recsys-paper} satisfy our requirements.
We test our ideas with two different types of recommender systems, 
namely collaborative filtering and deep neural network based.
Section~\ref{als-vs-dl-experiment} presents more details of our experiments
with different recommender systems.

Algorithm~\ref{alg:collection-recsys} presents the entire algorithm of our system. 
The function call \texttt{TrainItemRecSys()} represents the step of learning representations.
The size of the embeddings in our experiments is set to 100 for both items and users.

\subsection{Creating and recommending collections} 
\label{creating-and-recommending-collections}

Collection creation and recommendation 
operate on a per user level ($u \in U$).
Within the \texttt{foreach} loop,
it first generates top \texttt{numRecItems} item recommendations by calling
\texttt{GetTopNItems()}. 
The function takes the specific user embedding $u^{Emb}$, 
item embeddings $I^{Emb}$ and \texttt{numRecItems}.
We optimize the operation by indexing the item embeddings and doing an
approximate nearest neighbour lookup \cite{annoy}.
The value of \texttt{numRecItems} depends on the use case;
specifically, it depends on
the number of collections to be created and the number of items per collection.
For instance, in our case, we set it to 1000 for
creating up to 5 playlists (collections) each with 30 songs (items).

\vspace{0.3cm}
\noindent\begin{minipage}[htb]{.645\linewidth}
\begin{algorithm}[H]
\caption{Collection recommender system}\label{alg:collection-recsys}
\SetAlgoLined
\KwIn{User-item interaction data $D \gets \{(u, i, r_{ui}): \forall u\in U$ and $\forall i\in I\}$}
\KwOut{Collections $C^u$ for each user $u$}
\texttt{\\}
$U^\text{Emb}, I^\text{Emb} \gets \text{TrainItemRecSys}(D)$\\
\ForEach{$u \in U$}{
    $\text{recItems}^\text{Emb} \gets \text{GetTopNItems}(u^\text{Emb}, I^\text{Emb}, \text{numRecItems})$\\
    $\text{recItems}^\text{ReducedEmb} \gets \text{UMAP}(\text{recItems}^\text{Emb}, \text{reducedDimSize})$\\
    $G^u \gets \text{HDBSCAN}(\text{recItems}^\text{ReducedEmb})$\\
    $C^u \gets \phi$\\
    \For{$G_{k}^u$ in $G^u$}{
        $\{(i_{k}^u, r_k^u)\} \gets \text{GetTopRelevantItems}(G_{k}^u, \text{numItemsPerCluster})$\\
        $R_{k}^u \gets \text{GetClusterRating}(\{(i_{k}^u, r_k^u)\})$\\
        $\text{title}_{k}^u \gets \text{GetClusterTitle}(\{i_{k}^u\}, \text{Metadata})$\\  
        $C^u_k \gets (\text{title}_k^u, R_k^u, \{(i_{k}^u, r_k^u)\})$\\
        $C^u \gets C^u \cup \{C^u_k$\}\\
    }
    $C^u \gets \text{GetTopNClustersByRating}(C^u, N)$\\
}
\end{algorithm}
\end{minipage}
\hspace{.05\linewidth}
\begin{minipage}[htb]{.295\linewidth}
  \raggedright
  \includegraphics[width=0.85\linewidth]{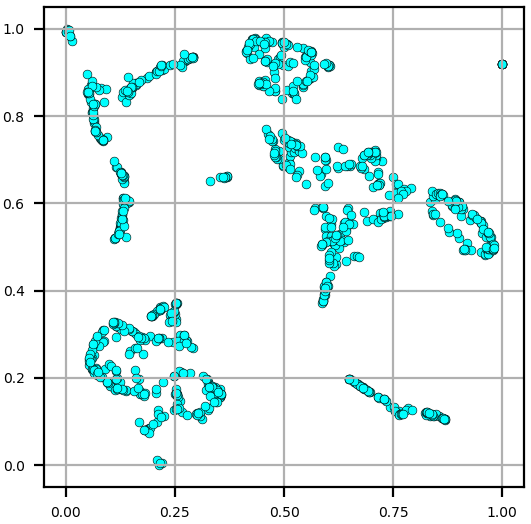}
  \captionof{figure}{Output after UMAP step} \label{fig:umap}
  \vspace{0.4cm}
  \includegraphics[width=\linewidth]{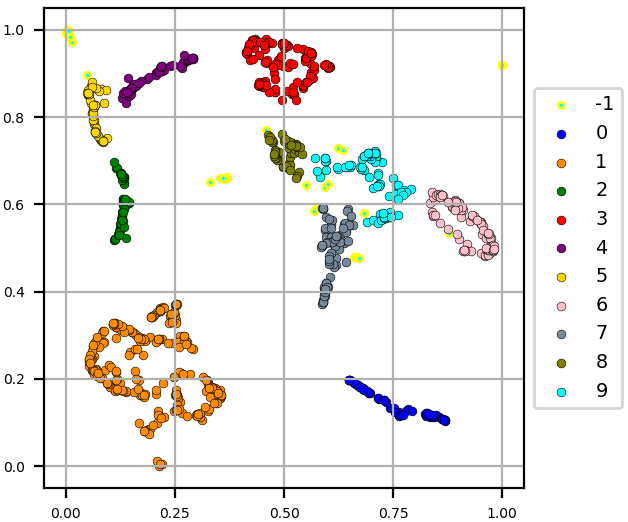}
  \captionof{figure}{Output after HDBSCAN step. Labels 0-9 are valid clusters and -1 represents noisy points.} \label{fig:hdbscan}
\end{minipage}
\vspace{0.3cm}

In the next step, we reduce the dimensionality of the embeddings of \texttt{recItems}
from 100 to 3 (\texttt{reducedDimSize}).
We use the Uniform Manifold Approximation and Projection (UMAP) \cite{umap-paper} algorithm,
which is a non-linear technique of dimensionality reduction.
Next, we use
the Hierarchical Density-Based Spatial Clustering of Applications with Noise 
(HDBSCAN) \cite{hdbscan-paper} algorithm to create clusters of \texttt{recItems} 
in the reduced dimension space, which returns groups of items $G^u$ for the current user.
We discuss our choice of the combination of UMAP and HDBSCAN 
for creating clusters in Section~\ref{sec:experiments}.

In the nested \texttt{for} loop, we iterate over groups $G^u$.
A group may contain items more than we require, and we need to carefully select 
the desired number of items from it.
The function
\texttt{GetTopRelevantItems()} is called for each group $G_k^u$ to 
select \texttt{numItemsPerCluster} items.
For our case, the value of \texttt{numItemsPerCluster} is 30.
Here, we have to select the items that are 
most relevant to the user at the same time they are
most representative of the cluster.
We use the following heuristic: we first select 
the most relevant item (item with highest rating $r^u$, 
i.e., highest user-to-item similarity) in the group;
and then, we select the rest of the \texttt{(numItemsPerCluster - 1)} items 
that are most similar to the first selected item based on item-to-item 
similarity.
This results in a collection that has items relevant to the user 
as well as maintains the notion of similarity within a collection.
Finally, the function returns the items with their ratings $\{(i_{k}^u, r_k^u)\}$, 
which will be used to create the collection $C_k^u$.
Note that the groups that have items less than \texttt{numItemsPerCluster} 
are ignored (not shown in the algorithm).

\texttt{GetClusterRating()} computes the rating of the collection $C_k^u$. 
It does so by averaging the ratings of the individual items.
The title of the collection is determined by calling \texttt{GetClusterTitle()} with
its arguments as items and the \texttt{Metadata} of the items.
The definition of this function is specific to the domain in consideration.
In our case, we select the top three artists (by frequency) present in a collection.
This is a very simple method, however,
because of the familiarity of the users with the artists,
it turns out to be very effective.
Showing the top three artists conveys multiple properties of a collection.
First and foremost, it tells that it is a collection (mix).
Second, it tells the language of the collection. Finally, 
it communicates the genre and the type of songs in the collection.

We collect all the collections in set $C^u$, and in the end, 
we call \texttt{GetTopNClustersByRating()} to select the most relevant
$N$ collections 
(by cluster rating $R_k^u$) 
for the user.

\section{Experiments and evaluation}
\label{sec:experiments}

We conduct our experiments in the domain of music recommendation,
within a music streaming service.
This service caters to a diverse population with its consumption spanning 
over 10 international
and 31 regional languages.
To perform our experiments, we took
a subset of our data, containing 
43 million users, 
5.1 million songs, and
3.8 billion
interaction points between them to learn user-item representations.
For training representations with side information (refer to Section~\ref{als-vs-dl-experiment}), 
we additionally took item attributes such as artist, language and year of release.

To measure the quality of the collections (playlists) created and recommended, 
we define 
multiple offline and online metrics.
We use offline metrics to develop the algorithm and tune the hyperparameters, 
and we use online metrics to measure the user engagement.

We measure the following offline metrics: 
1) The average number of collections generated per user.
2) The percentage of collections containing language noise. 
A collection is said to contain language noise if more than 20 \% of 
its items are of different languages.
Though a lower language noise is desired, 
manual inspection revealed that few pairs of languages frequently come together, 
as songs of these languages are commonly streamed together in a session.
This sort of noise is good to have, to a certain extent, 
because we want to create collections as per the
user consumption patterns.
3) The percentage of collections with intra-cluster noise. 
A collection is said to contain intra-cluster noise 
if it contains more than 20 \% anomalies. 
An item in a collection is an anomaly if its item-to-item 
similarity from the most relevant item in the collection 
(refer to Section~\ref{creating-and-recommending-collections}) 
is lower than a threshold of 0.2. Having a lower intra-cluster noise is desirable.
4) The percentage of users observing overlap of artists across the titles of their collections.
In an ideal scenario, there should not be an overlap, but it is also 
found that certain overlap is good. 
For example, two collections may have common artists with 
their songs spanning multiple eras,
i.e., different themes from the same artists.
Because of this reason, the overlap is reported only when four or more collections have common artists in their titles.
5) Artist relevancy is the average percentage of songs in the collections 
with artists that the user has listened to in the past. 
High artist relevancy suggests that the songs in the collections 
are mostly from artists familiar to the user; conversely, 
it also suggests that exploration of new artists is limited.
6) Average rank is a measure of relevancy which 
represents the average position given by the recommender 
system to the items which make the collections. 
In a list of recommended items, more relevant items are present at the top, 
so, a lower average rank suggests that more relevant items are contained in the collections.

A careful reader would notice that some of these metrics are domain-specific and
metadata-dependent. However, all of these could be easily 
ported to a new domain.
Also, note that these metrics are not precise and just give us a coarse 
measure of the performance.
Thus, we also relied on manual inspections for a set of diverse users. 
It helped us significantly in generating insights and improving the algorithm.

The following sections present
ablation and variation experiments over 10k randomly 
selected users.
In an experiment,
we remove or change a single component
and measure the aforementioned metrics to check the efficacy.
For each experiment, we tune the hyperparameters and pick the best performing
configuration.
Given the unsupervised nature of our approach and the lack of 
ground truth to test the resulting collections, these metrics 
allowed us to experiment quickly and 
achieve our final algorithm.

\begin{table}[H]
\small
\begin{tabular}{p{0.17\textwidth} r r r r r r r r}
\hline
\textit{Representation learning}                                                                                                 & \textit{ALS}                                                        & \textit{ALS}                                                & \textit{ALS}                                               & \textit{ALS}                                                      & \textit{ALS}                                                      & \textit{ALS}                                               & \textit{ALS}                                                     & \textit{DNN}                                                        \\ 
\textit{Dim. reduction}                                                                              & \textit{UMAP}           & \textit{None}                                              & \textit{PCA}                                                     & \textit{t-SNE}                                    & \textit{UMAP}                                              & \textit{UMAP}                                              & \textit{UMAP}                                                     & \textit{UMAP}                                                                                                 \\ 
\textit{Clustering method}                                                                                       & \textit{HDBSCAN}                  & \textit{HDBSCAN}                                           & \textit{HDBSCAN}                                                 & \textit{HDBSCAN}                                               & \textit{k-means}                                           & \textit{Spectral}                                            & \textit{DBSCAN}                                                 & \textit{HDBSCAN}                                                                                           \\ \hline
\textit{\begin{tabular}[c]{@{}l@{}}average \# of collections \\ generated per user\end{tabular}} & 4.5         
& 2.5                                                      
& 3.2                                                             
& 4.3     
& 4.8                                             
& 4.5                                                             
& 4.6  
& 4.1   \\ \hline
\textit{language noise}                                                                  & \begin{tabular}[c]{@{}r@{}}2.7 \% \end{tabular}            &
\begin{tabular}[c]{@{}r@{}}4.9 \% \end{tabular}   &
\begin{tabular}[c]{@{}r@{}}5.1 \% \end{tabular}         & \begin{tabular}[c]{@{}r@{}}3.6 \% \end{tabular} &
\begin{tabular}[c]{@{}r@{}}2.9 \% \end{tabular}   & 
\begin{tabular}[c]{@{}r@{}}3.3 \% \end{tabular}          & \begin{tabular}[c]{@{}r@{}}2.3 \% \end{tabular}   &
\begin{tabular}[c]{@{}r@{}}3.6 \% \end{tabular}   \\ \hline
\textit{\begin{tabular}[c]{@{}l@{}}intra-cluster noise\end{tabular}}        &
\begin{tabular}[c]{@{}r@{}}4.2 \% \end{tabular}            &
\begin{tabular}[c]{@{}r@{}}4.1 \% \end{tabular}    &
\begin{tabular}[c]{@{}r@{}}10.3 \% \end{tabular}          & \begin{tabular}[c]{@{}r@{}}8.2 \% \end{tabular}         &
\begin{tabular}[c]{@{}r@{}}6.3 \% \end{tabular}   &
\begin{tabular}[c]{@{}r@{}}5.3 \% \end{tabular}          & \begin{tabular}[c]{@{}r@{}}3.5 \% \end{tabular}     &
\begin{tabular}[c]{@{}r@{}}1.5 \% \end{tabular}     \\ \hline
\textit{\begin{tabular}[c]{@{}l@{}}\% users with title overlap \\ in 4 or more collections\end{tabular}} & 
\begin{tabular}[c]{@{}r@{}}7.7 \% \end{tabular}    & 
\begin{tabular}[c]{@{}r@{}}0.0 \% \end{tabular}      & 
\begin{tabular}[c]{@{}r@{}}1.5 \% \end{tabular}          & \begin{tabular}[c]{@{}r@{}}8.3 \% \end{tabular}     &
\begin{tabular}[c]{@{}r@{}}12.7 \% \end{tabular}  &
\begin{tabular}[c]{@{}r@{}}9.4 \% \end{tabular}           & \begin{tabular}[c]{@{}r@{}}8.7 \% \end{tabular}    &
\begin{tabular}[c]{@{}r@{}}3.9 \% \end{tabular}    \\ \hline

\textit{artist relevancy}                                                                  & \begin{tabular}[c]{@{}r@{}}61.8 \% \end{tabular}            &
\begin{tabular}[c]{@{}r@{}}55.9 \% \end{tabular}   &
\begin{tabular}[c]{@{}r@{}}58.2 \% \end{tabular}         & \begin{tabular}[c]{@{}r@{}}60.7 \% \end{tabular} &
\begin{tabular}[c]{@{}r@{}}61.6 \% \end{tabular}   & 
\begin{tabular}[c]{@{}r@{}}61.3 \% \end{tabular}          & \begin{tabular}[c]{@{}r@{}}61.6 \% \end{tabular}   &
\begin{tabular}[c]{@{}r@{}}60.5 \% \end{tabular}   \\ \hline

\textit{average rank}                                                                  & \begin{tabular}[c]{@{}r@{}}173.7 \end{tabular}            &
\begin{tabular}[c]{@{}r@{}}208.4 \end{tabular}   &
\begin{tabular}[c]{@{}r@{}}197.2 \end{tabular}         & \begin{tabular}[c]{@{}r@{}}175.6 \end{tabular} &
\begin{tabular}[c]{@{}r@{}} 174.0 \end{tabular}   & 
\begin{tabular}[c]{@{}r@{}} 173.8 \end{tabular}          & \begin{tabular}[c]{@{}r@{}} 174.1 \end{tabular}   &
\begin{tabular}[c]{@{}r@{}} 184.8 \end{tabular}   \\ \hline
\end{tabular}
\vspace{0.2cm}
\caption{Offline experiments. Second column onwards represent a specific experiment; 
e.g., 
the fourth column represents the experiment with ALS, PCA and HDBSCAN as the
representation learning, dimensionality reduction and clustering methods, respectively.
The rows represent different metrics.
}  
\label{table:offline-metrics}
\end{table}
\subsection{Need and choice of dimensionality reduction algorithm}  
\label{dim-red-experiment}

Dimensionality reduction transforms data from a high-dimensional space
to a lower dimension. In our case, we use it to transform
item embeddings from a vector of 100 dimensions to
3 dimensions.
It reduces the sparsity of our data, 
overcoming the curse of dimensionality \cite{clustering-challenges-paper},
and the next step of clustering is able to
create better collections.
We conduct four experiments:
one without dimensionality reduction and
the rest with three different algorithms, namely 
UMAP \cite{umap-paper}, PCA \cite{pca-paper} and t-SNE \cite{tsne-paper}.
Columns \{\textit{ALS, None, HDBSCAN}\}, \{\textit{ALS, UMAP, HDBSCAN}\}, \{\textit{ALS, PCA, HDBSCAN}\} and \{\textit{ALS, t-SNE, HDBSCAN}\} of
Table~\ref{table:offline-metrics} present the metrics for
these four experiments, respectively.

We draw the following conclusions:
1) Dimensionality reduction
leads to an increase in the number of collections created no matter which
algorithm is used.
2) Non-linear approaches to dimensionality reduction (UMAP and t-SNE) 
perform significantly better than the linear approach of PCA. 
Among those, UMAP performs better in our case.

Note that we do not show the experiments for choosing the number of dimensions 
of the reduced space.
Nevertheless, we found 3 to be the optimal value.

\subsection{Choice of clustering algorithm}  \label{clustering-algos-experiment}

Figure~\ref{fig:umap} shows the data points 
after reducing to a 2-dimensional space using UMAP. 
We can 
see that the clusters can potentially be of arbitrary shapes and with variable 
densities.

We try four different clustering algorithms, namely 
k-means clustering \cite{kmeans-clustering-paper},
spectral clustering \cite{spectral-clustering-tutorial-paper},
DBSCAN \cite{dbscan-paper} and
HDBSCAN \cite{hdbscan-paper}.
k-means clustering can only find convex/spherical clusters in a dataset, 
while others can find clusters of arbitrary shapes. 
Unlike k-means and spectral clustering methods, DBSCAN and HDBSCAN 
use density-based clustering and
do not require the number of clusters to be specified.
HDBSCAN can additionally identify variable density clusters.
Figure~\ref{fig:hdbscan} shows the application of HDBSCAN on 
the output of UMAP.

The results of these four experiments are shown in the
columns \{\textit{ALS, UMAP, k-means}\}, \{\textit{ALS, UMAP, Spectral}\}, \{\textit{ALS, UMAP, DBSCAN}\} and \{\textit{ALS, UMAP, HDBSCAN}\} of Table~\ref{table:offline-metrics}.
We make the following conclusions:
1) DBSCAN and HDBSCAN have lower noise and title overlaps in 
comparison to k-means and spectral clustering methods, as desired.
2) DBSCAN performs marginally better than HDBSCAN 
in the average number of collections generated per user, 
language noise and intra-cluster noise, 
whereas HDBSCAN does so in the remaining metrics.

\subsection{Varying representation providers}  \label{als-vs-dl-experiment}
We try two different recommendation systems to learn user-item
representations, namely a collaborative filtering (CF) model   
and a two-tower deep neural network model (DNN).
The CF model
is a parallelized version of matrix factorization  
with alternating least squares optimizer (ALS)
\cite{als-wr-paper}.
The DNN model \cite{deep-learning-recsys-paper} consists of 
a user tower and an item tower. 
The former contains a single dense layer of 100 units, 
while the latter contains a dense layer of 110 units 
(the additional 10 units are for item metadata,
namely artists, language and release year) 
followed by the dense layer of 100 units.
We train both the recommendation models to learn user-item embeddings. 
In case of ALS, these are the embedding values in the user and item matrices, 
whereas in DNN, these are the activations of the penultimate 
layers of the model (i.e., the final layer of each tower).

Columns \{\textit{DNN, UMAP, HDBSCAN}\} and \{\textit{ALS, UMAP, HDBSCAN}\}  of Table~\ref{table:offline-metrics} report the results for the two kinds of representations.
It is interesting to note that the same configuration of our algorithm
works for both the models and produces similar metrics, 
we just marginally tuned the hyperparameters of HDBSCAN.
Some of the differences that we see are due to the fact that
the 
representations learnt
by DNN model are much more accurate --- it uses the information about the artists
of the items --- and 
it leads to a lesser overlap noise.

\subsection{Online experiments} \label{online-experiments}
Based on the above experiments, we select UMAP
as our dimensionality reduction technique 
and HDBSCAN as our clustering method. 
Specifically, we finalize the following hyperparameters:
for UMAP\footnotemark
\footnotetext{UMAP Python API: \href{https://umap-learn.readthedocs.io/en/latest/api.html}{\texttt{https://umap-learn.readthedocs.io/en/latest/api.html}} (Visited on April 29, 2021)}, $n\_neighbors=15$, $n\_components=3$, $min\_dist=0.0$,
and for HDBSCAN\footnotemark
\footnotetext{HDBSCAN Python API: \href{https://hdbscan.readthedocs.io/en/latest/api.html}{\texttt{https://hdbscan.readthedocs.io/en/latest/api.html}} (Visited on April 29, 2021)}, $min\_samples=8$, $min\_cluster\_size=30$.
With our algorithm fixed, we conduct online experiments to measure user engagement.
The recommendation system for learning user-item representations
is ALS.
We compare the results between the settings of recommending individual items versus
recommending collections of items. 
We do so through an A/B experiment over randomly distributed, 
equally sized userbases.
We show item recommendations to userbase A and collection recommendations
to userbase B at the same top-most position on the homepage.
More concretely, in the former, we show the top 15 songs. 
In the latter,
we show the top 5 collections, where each collection can be clicked to see the 30 songs inside it.

The results are as follows:
1) On an average, the userbase A consumes 2.94 songs per day through item recommendation, 
whereas the userbase B consumes 6.83 songs per day through collection recommendation -- 
a 2.3x increase in recommendation-driven song consumption.
Here, the consumption of a song is defined as
at least 30 seconds stream of the song.
2) Of the total songs a user consumes on the music streaming service, 
16.7 \% of them are from item recommendation in setting A, 
while 36.5 \% of them
are from collection recommendation in setting B.
Thus, users start relying more on recommendations.
3) The users have an average click-through rate (CTR) of 
18.9 \% in setting A as compared to 15.2 \% 
in setting B. Here, CTR is defined as the ratio of 
the users who click on an entity (a tile on real estate) to 
the number of users who see it. 
A lower value of CTR in setting B is expected because we show 
only 5 entities (collections) in setting B in comparison to 
15 entities (items) in setting A.
4) Although the consumption driven by recommendation increased significantly, but
the total consumption, i.e., from recommendation as well as the other sources, 
only increased marginally by 0.07 songs per day per user.
It implies that users' consumption shifted to 
the recommendation source
from
the other sources such as user library, albums, and editorials.

The above results are extremely encouraging and highlight the 
significance of providing collections of relevant songs to our customers. 
It is evident that once a user selects a theme (collection), 
he/she tends to play the entire playlist, 
thus resulting in a more engaged user. More so, 
since collections are hierarchical in nature, we now recommend
a total of 150 songs across 5 distinct themes 
in contrast to 
only 15 songs individually, 
thus improving the utilization of the same real estate.
Essentially, it enables us to recommend more and diverse
content in an organized manner.

\section{Related Work}

We have categorized related literature
into four areas, namely 
the work that focuses on the idea and usability of collection recommendation,
the work on automatic collection recommendation,
the work on semi-automatic collection recommendation, and
finally, a somewhat related area of topic modeling and clustering.

\textbf{Concept and usability:}
Hansen and Golbeck conceptualized the idea of the collection recommender systems from the point of 
view of significance and usability \cite{hansen-golbeck-mix-paper, hansen-golbeck-framework-paper}.
The authors presented the need for collection recommendation 
instead of individual item recommendations 
and discussed the challenges and design decisions around
algorithm development, preference elicitation, 
evaluation techniques, and user interface.
They also discussed different types of collections 
and the notions of similarities between the items
to form a collection.
There are more related papers that conceptualized collection 
recommendations in various settings 
\cite{recsys-new-perspectives-paper, curator-paper}.
However, none of the aforementioned works 
\cite{hansen-golbeck-mix-paper, hansen-golbeck-framework-paper, recsys-new-perspectives-paper, curator-paper} 
provides a concrete implementation and tests the ideas on a real-world problem.
In contrast, we present the design and implementation of a system 
that can automatically recommend collections
in a setting of music recommendation.

\textbf{Automatic collection recommendation:}
A recent work from Netflix \cite{netflix-collections-paper} presented a method to recommend 
\emph{existing} collections. That is, given a set of collections already available,
the method predicts the ratings of the collections ($R^u$) as well as the ratings of 
the items ($r^u$) within a collection w.r.t. a user ($u$).
This enables them to decide which collections to recommend and the order of the items
within a collection.
Our work also falls in the category of automatic collection recommendation, however,
we do not make the assumption of the availability of collections in 
advance. 
In fact, our algorithm creates the collections as well.

\textbf{Semi-automatic collection recommendation:}
Semi-automatic collection recommendation systems require user input
to recommend collections.
The input could be the constraints on a collection, 
e.g., the total time of a playlist  \cite{packages-paper}
or the total cost of a trip consisting of places to visit in one or more cities 
\cite{comprec-trip-paper};
it could also be the \emph{seed} songs
in case of music, which are used by the
system to create playlists using item-item similarity based on item embeddings or metadata 
\cite{rush-paper, packages-paper, toprecs-paper, card-paper, comprec-trip-paper}.
Clearly, our work is more general than this class of collection recommenders, as 
we do not require any manual input from a user in the process of recommending collections.

\textbf{Topic modeling and clustering:}
Although there is no notion of users and recommendation in the problem of 
topic modeling (natural language processing), 
 it comes close to the idea of clustering.
Top2Vec \cite{top2vec-paper} 
uses doc2vec \cite{doc2vec-paper} to generate document embeddings,
 UMAP to reduce their dimensionality, 
and finally, HDBSCAN to create clusters in the smaller dimension space.
We too use UMAP to reduce embeddings dimensions and HDBSCAN to cluster, however, 
we differ in the way we detect the top items in 
a cluster. Moreover, we predict the ratings of the clusters and 
that of the items within, which is not the case with topic modeling.
In fact, there is no notion of users in topic modeling.

ClusterExplorer \cite{clusterexplorer-paper} uses the idea of clustering of 
item embeddings learnt through a matrix factorization (user-item interaction) 
recommender system. 
The objective of the work is to provide ``related recommendations''
of items and topics based on a selected item. 
It uses k-means clustering along with DBSCAN 
to create topics.
In contrast, our goal is to recommend personalized collections whereas their
objective is to predict
similar items and topics without any personalization.

\section{Conclusion and future work}
%

We presented a \emph{collection recommender system} that can automatically
create and recommend collections of items.
Unlike regular recommender systems, which output
top-N relevant items, a collection recommender system
outputs collections of items such that the items in the collections are relevant to a user, 
and items within a collection follow a specific theme.
Our system is built on top of the user-item representations learnt by 
item recommender systems.
We employed dimensionality reduction and clustering techniques 
along with intuitive heuristics 
to create collections with their ratings and titles.
We tested these ideas in a real-world setting of music recommendation, 
within a popular music streaming service.
We found that there is a 2.3x increase in recommendation-driven consumption
when recommending collections over items.
Moreover, it also resulted in effective utilization of
real estate and lead to recommending a more and 
diverse set of items in an organized manner.

We believe our work is just a beginning in the direction of
perfecting 
collection recommendation.
There is a lot of ground to be explored, including 
applications in other domains, 
building better methods,
experimenting with different types of datasets,
devising new metrics for measuring the performance of collections,
and taking user feedback for improving the 
algorithm.
We hope that our work brings attention to this problem and
induces more studies around this problem.

\bibliographystyle{ACM-Reference-Format}
\bibliography{main}

\end{document}